\documentclass[conference]{IEEEtran}
\IEEEoverridecommandlockouts
\usepackage{cite}
\usepackage{amsmath,amssymb,amsfonts}
\usepackage{algorithmic}
\usepackage{graphicx}
\usepackage{textcomp}
\usepackage{hyperref}
 \usepackage{mathtools}
 \usepackage{commath}

\usepackage{subcaption}
\usepackage{xcolor}
\def\BibTeX{{\rm B\kern-.05em{\sc i\kern-.025em b}\kern-.08em
    T\kern-.1667em\lower.7ex\hbox{E}\kern-.125emX}}

\makeatletter
\newcommand{\linebreakand}{%
  \end{@IEEEauthorhalign}
  \hfill\mbox{}\par
  \mbox{}\hfill\begin{@IEEEauthorhalign}
}
\makeatother
\begin{document}

\title{ART: A Graph-based Framework for Investigating Illicit Activity in Monero via Address-Ring-Transaction Structures}

\author{\IEEEauthorblockN{Andrea Venturi,}
\IEEEauthorblockA{\textit{Digital Security,} \\
\textit{Vicomtech Foundation}\\
Donostia, Spain \\
aventuri@vicomtech.org}
\and
\IEEEauthorblockN{Imanol Jerico-Yoldi}
\IEEEauthorblockA{\textit{Digital Security,}\\
\textit{Vicomtech Foundation}\\
Donostia, Spain\\
ijerico@vicomtech.org}
\and
\IEEEauthorblockN{Francesco Zola}
\IEEEauthorblockA{\textit{Digital Security,}\\
\textit{Vicomtech Foundation}\\
Donostia, Spain\\
fzola@vicomtech.org}
\and
\IEEEauthorblockN{Raul Orduna}
\IEEEauthorblockA{\textit{Digital Security,}\\
\textit{Vicomtech Foundation}\\
Donostia, Spain\\
rorduna@vicomtech.org}}

\maketitle

\begin{abstract}
As Law Enforcement Agencies advance in cryptocurrency forensics, criminal actors aiming to conceal illicit fund movements increasingly turn to “mixin” services or privacy-based cryptocurrencies. Monero stands out as a leading choice due to its strong privacy preserving and untraceability properties, making conventional blockchain analysis ineffective. Understanding the behavior and operational patterns of criminal actors within Monero is therefore challenging and it is essential to support future investigative strategies and disrupt illicit activities. In this work, we propose a case study in which we leverage a novel graph-based methodology to extract structural and temporal patterns from Monero transactions linked to already discovered criminal activities. By building Address-Ring-Transaction graphs from flagged transactions, we extract structural and temporal features and use them to train Machine Learning models capable of detecting similar behavioral patterns that could highlight criminal modus operandi. This represents a first partial step toward developing analytical tools that support investigative efforts in privacy-preserving blockchain ecosystems. 
\end{abstract}


\section{Introduction}

The decentralized and pseudo-anonymous nature of cryptocurrencies has made them a favored tool for organized criminal groups seeking to conceal the flow of illicit funds. In response, Law Enforcement Agencies (LEAs) have increasingly invested in blockchain forensics, developing techniques to analyze and disrupt the anonymity level provided by transparent blockchains like Bitcoin. These advances have driven criminal actors to adopt further obfuscation strategies, including the use of crypto-mixers [3] and a growing shift toward privacy-focused cryptocurrencies.
Among these, Monero has emerged as one of the most widely used and technically advanced privacy coins. Launched in April 2014, Monero is specifically designed to provide strong privacy guarantees through its RingCT protocol, which ensures both untraceability and unlinkability of transactions \cite{sun2017ringct}. Its adoption has been documented in several criminal contexts. For instance, the January 2024 Chainalysis report highlights the recent shift of Child Sexual Abuse Material (CSAM) vendors towards instant exchanges offering Monero \cite{Chainalysis}. Similarly, Monero has been linked to high-profile cybercrime operations, including the Lazarus Group’s to launder proceeds from the WannaCry 2.0 ransomware campaign \cite{bax2021}, and LockBit Group for gathering the ransom after their attacks \cite{walter2022lockbit}.
This work takes an initial step toward understanding and characterizing the modus operandi of criminals on the Monero blockchain. While previous research has primarily focused on developing traceability methods or identifying security vulnerabilities to compromise Monero’s cryptographic protections (e.g., \cite{moser2017empirical,cremers2023holistic}), our approach shifts the focus toward behavioral analysis. We introduce a graph-based method that models Monero’s transaction ecosystem through the Address-Ring-Transaction (ART) graph, a novel representation specifically designed to capture unique properties of Monero’s privacy-preserving protocol. This representation enables the extraction of structural and temporal features from transactions associated with specific behaviors, which can then be used to train Machine Learning (ML) classifiers capable of identifying other transactions exhibiting similar patterns within the blockchain. We validate our methodology through a case study based on real Monero transactions linked to illicit activities, demonstrating the potential of behavioral graph-based analysis as a complementary means for investigating privacy-oriented cryptocurrencies.
By focusing on behavioral patterns rather than protocol-level vulnerabilities, this approach uncovers transactional regularities that can guide future detection strategies and pave the way for advanced investigative tools in privacy-preserving blockchain systems

\section{Background and Problem Statement}\label{sec:background}
Monero’s privacy model is primarily based on the Ring Confidential Transactions (RingCT) protocol \cite{sun2017ringct}, which combines three fundamental privacy-enhancing technologies: ring signatures, stealth addresses, and confidential transactions. Ring signatures enable a sender to obscure their identity within a group of potential signers. In Monero, every transaction input is mixed with decoy inputs (also called mixins) from the blockchain, making it computationally infeasible to determine which input is the actual spender. Stealth addresses enable recipients to receive funds via one-time addresses generated from their public key, ensuring that transactions cannot be linked back to their actual address. Confidential transactions use Pedersen commitments to conceal transaction amounts. This prevents third parties from viewing or analyzing transferred values while still allowing network participants to verify transaction validity.
Together, these technologies make Monero a preferred tool by individuals who aim to obscure their financial movements, including actors involved in illicit activities, such as money laundering, ransomware payments and darknet market transactions. As a consequence, understanding and modeling the behavior and modus operandi of criminals leveraging Monero is thus a critical task. Our goal in this work is to provide a first analytical method that could support LEAs in detecting common patterns exhibited by criminals when performing obfuscation operations in Monero.

\section{Related Work}\label{subsec:related}
A significant portion of the existing literature on Monero focuses on transaction traceability by developing heuristic methods aimed at identifying the true input among ring signatures, rather than examining potential criminal modus operandi. Seminal works in this area, such as \cite{moser2017empirical} and \cite{zola2025topological}, introduced heuristic-based attacks capable of significantly narrowing the set of potential senders, thereby undermining anonymity. However, these techniques were only effective against early versions of the Monero protocol and are no longer applicable. More recent approaches have been proposed \cite{Chainalysis}, but their effectiveness is limited, and they typically require stringent conditions and considerable computational resources.
In contrast, our work addresses the problem from a different angle, focusing on behavioral analysis within the Monero blockchain. This offers a novel and complementary perspective to the current state of the art. While our graph-based methodology is inspired by prior research on behavioral analysis (e.g., \cite{bax2021}), to the best of our knowledge, this is the first time such an approach has been applied to Monero, and previous studies have generally focused on traditional, transparent cryptocurrencies (e.g., \cite{colladon2017using,tariq2023topology,usman2023intelligent}).

\section{Behavior Detection Method}\label{sec:pattern}
We introduce a novel graph-based method to uncover behavioral patterns and identify specific—potentially illicit—modus operandi within the Monero blockchain. Unlike transparent cryptocurrencies such as Bitcoin, Monero’s strong privacy guarantees make direct observation of user activity infeasible, posing a fundamental challenge to traditional blockchain analysis. Our approach addresses this limitation by modeling the Monero network through a new graph representation that captures both its structural and temporal dynamics.
Our approach is divided into four main phases which are detailed below.

\subsection{Data Acquisition}
The main objective of the proposed methodology is to train binary classifiers capable of detecting transaction patterns that resemble a given modus operandi inside the Monero blockchain. A crucial step in this process is the acquisition of sufficient labeled data that accurately represents the behavioral characteristics of the target activity. Due to the privacy mechanisms implemented in Monero—such as one-time stealth addresses and ring signatures—traditional address-based tracing is not feasible. Therefore, our analysis focuses on transactions themselves, which constitute the most accessible and reliable source of information within Monero’s privacy framework.
By obtaining a set of transactions labeled according to the modus operandi of interest, we can observe and reconstruct the associated transaction flows, enabling the identification of distinct behavioral patterns linked to specific entities. To achieve this goal, we design a graph-based framework capable of representing the structure and evolution of Monero transactions.

\begin{figure}
    \centering
    \includegraphics[width=\linewidth]{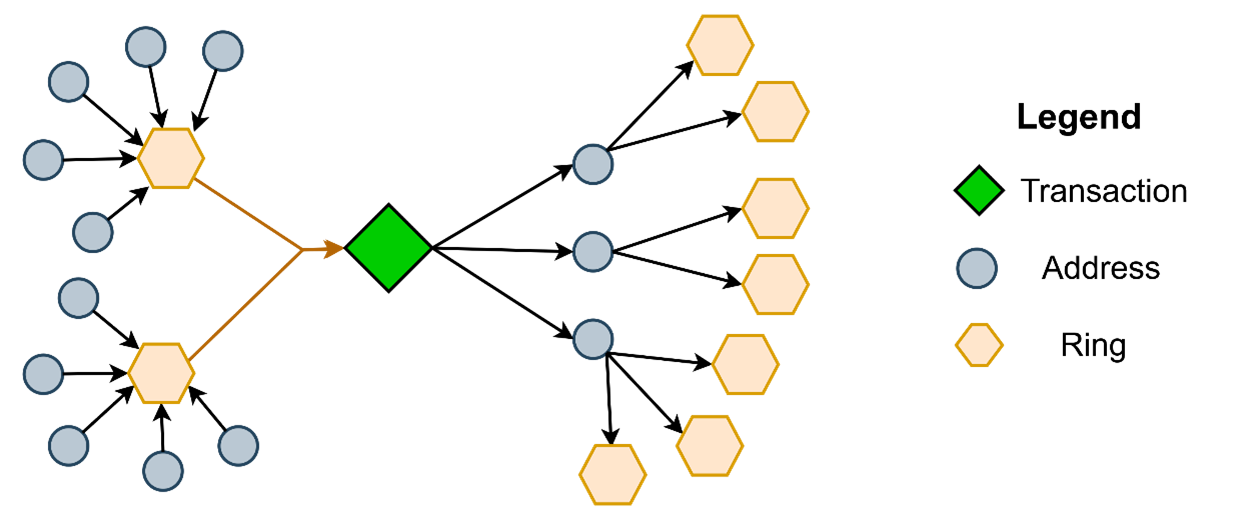}
    \caption{Example of ART-graph}
    \label{fig:enter-label}
\end{figure}

\subsection{ART-graph}
We propose the Address-Ring-Transaction (ART) graph, an adaptation of the conventional address-transaction graph widely used in blockchain analytics \cite{zola2025topological}. While previous graph-based studies rely on explicit transaction links (e.g., \cite{cremers2023holistic}), the ART-graph is specifically designed to reflect Monero’s privacy-preserving mechanisms. It includes three distinct types of nodes - transactions, addresses, and rings - to accurately represent the relationships introduced by Monero’s ring-signature protocol. 
Figure 1 provides a schematic example of an ART-graph for a single transaction. The green square node denotes the transaction itself, connected on the left to its two input rings (orange hexagons). Each ring aggregates several stealth addresses (gray circles), which are one-time addresses derived from outputs of previous transactions. Because these addresses may appear in multiple rings, they form indirect connections between otherwise unlinkable transactions. Importantly, within each ring only one member corresponds to the true input, whereas the others serve as decoys, thereby obfuscating the real flow of funds. For these reasons, the conventional definition of user behavior - typically based on the activity of a persistent cryptocurrency address – does not apply here. Hence, our analysis builds the ART-graph from transactions rather than from addresses, treating each transaction as the primary entity from which the subsequent analysis is derived. It is also important to remark that the ART-graph can be expanded in a hierarchical fashion. In particular, we refer to n-hop ART-graphs as cases in which the graph is iteratively expanded in the output direction, progressively including transactions that are farther from the one under analysis. This hierarchical expansion allows for the modeling of increasingly broader behavioral contexts around each transaction.

\subsection{Feature extraction}
The feature extraction phase starts by building an n-hop ART-graph for each transaction in the dataset. From each resulting graph, we extract a set of structural and temporal features which are then associated with the corresponding seed transaction. These features are designed to capture behavioral patterns that can characterize a specific modus operandi.
Referring to the transaction under analysis as TX\_A, the structural and temporal features we extract from its n-hop ART-graph can be organized into different blocks:
\begin{enumerate}

\item \textit{0-hop features:} these features are extracted directly from TX\_A, without constructing its ART-graph. 0-hop features include the number of rings, ring size, number of outputs, transaction fee and the count of unique ring members across all inputs. Temporal features in this block consist of the average (and standard deviation), minimum, maximum and median of the time differences between the timestamp of TX\_A and the timestamps of the transactions that produced the ring members.
\item \textit{i-hop features:} these features summarize the transactions located at the i-th hop from TX\_A within the ART-graph (where i ranges from 1 to n). For this set of transactions, we compute the minimum, maximum, average (std. dev.), and median of the number of rings per transaction and of the ring sizes. In addition, we calculate the same statistics also for the time differences between the timestamps of i-hop transactions and the timestamp of TX\_A.
    
\end{enumerate}
We remark that larger value of n broadens the topological and temporal contexts considered around TX\_A, as additional layers of neighboring transactions contribute new feature blocks. However, this also entails higher computational complexity and processing time, as the ART graph expands exponentially with each additional hop.

\subsection{Classification}
The extracted features can then be employed to train Machine Learning (ML) models aimed at recognizing transactions that exhibit behavioral similarities to those previously observed. By learning from the structural and temporal patterns encoded in the feature space, these models can generalize beyond the analyzed cases and detect other instances reflecting comparable operational characteristics. This enables the automatic identification of recurring behavioral signatures within the blockchain, supporting both investigative analyses and the development of data-driven tools for monitoring suspicious activity in privacy-focused cryptocurrencies.
Considering as positive the transactions labeled according to the modus operandi under investigation,  we propose to evaluate model performance using standard binary classification metrics, namely Precision, Recall, and F1-score, which are defined as follows:

\[
\text{Recall} = \frac{TP}{TP + FN}
\]

\[
\text{Precision} = \frac{TP}{TP + FP}
\]

\[
\text{F1-score} = \frac{2 \cdot \text{Precision} \cdot \text{Recall}}
{\text{Precision} + \text{Recall}}
\]

in which the terms True Positives (TP), False Positives (FP), True Negatives (TN), and False Negatives (FN) denote the correctly and incorrectly classified instances in each category. Precision measures the proportion of transactions identified as positives that are correctly classified, while recall quantifies the fraction of truly positive transactions that the model successfully detects. The F1-score, defined as the harmonic mean of precision and recall, provides a balanced view of detection performance.

\section{Case Study}\label{sec:data}
The case study presented in this section serves to validate the proposed methodology using real-world data from the Monero blockchain. In particular, we base our analysis on publicly available labeled data, which identifies specific transactions as part of the WannaCry 2.0 ransomware campaign, reportedly linked to the Lazarus Group \cite{bax2021}. According to these reports, the group is believed to have converted Bitcoin into Monero in August 2017. Subsequently, approximately three months later, they moved funds back into Bitcoin and Bitcoin Cash. 
In the following, we describe the dataset used in the case study, the features extracted, the implementation details, and the results obtained.

\subsection{Dataset and features}\label{subsec:data}
The report in \cite{bax2021} identifies a total of 19 Monero transactions associated with the operations of the Lazarus Group during the WannaCry 2.0 campaign. According to the authors, this tracing was made possible through the use of publicly available information, particularly leveraging data from the Shapeshift API – a cryptocurrency exchange service used by the Lazarus Group to convert funds from Bitcoin to Monero and subsequently back to Bitcoin Cash and Bitcoin. The reports divide the 19 transactions into three groups: 8 entering transactions (all on 03 August 2017), 3 consolidating transactions (17 August 2017) and 8 exiting transactions (02 November 2017). Notably, at the time these operations occurred, RingCT had already been implemented in Monero, further increasing the complexity of transaction analysis by concealing both the amounts and links between inputs and outputs.
All these transactions are labeled collectively as belonging to the positive class. To construct a binary dataset, we randomly sampled 150 additional transactions from the same three-month period as the positive samples to serve as negative examples. This process resulted in a suitable dataset composed of both positive and negative transactions, ready to be passed to the subsequent phases.
Once the dataset is defined, we proceed with the feature extraction phase. For each transaction in the dataset, we construct its corresponding 2-hop ART-graph. The graph construction process starts from the transaction under analysis. From this transaction, we identify its output addresses, connect these addresses to the rings in which they participate—whether as actual spenders or as mixins—and further link those rings back to their corresponding transactions. We build ART-graphs incrementally, expanding them up to two hops. As described before, each hop level adds transactions connected through output addresses and rings, with higher hops progressively including transactions farther from the initial labeled transaction. Finally, we derive the sets of structural and temporal features described in Section 4.3 to be used for training the machine learning classifier.

\subsection{Model implementation and training}\label{subsec:config}
After the feature extraction phase, we trained a supervised ML classifier to distinguish between transactions associated with illicit activity and regular ones. To address the imbalance between the two classes (19 positive transactions against 150 negative transactions), we employed the Synthetic Minority Oversampling Technique (SMOTE), which generates synthetic samples of the minority class to improve the model’s ability to recognize rare behaviors. 
For classification, we employed a Random Forest (RF) model, chosen for its robustness, capacity to capture nonlinear relationships, and suitability for heterogeneous numerical features. The dataset was divided into 80

\subsection{Results}
The RF classifier achieved encouraging results despite the limited dataset size (Table 1). Specifically, the model obtained a Precision of 1.000, Recall of 0.750, and an F1-score of 0.857 on the test set. These results suggest that the features extracted from the ART-graph effectively capture behavioral and structural patterns associated with the illicit activity attributed to the Lazarus Group. 
The perfect precision indicates that all transactions predicted as illicit were correctly classified, meaning the model produced no false positives in the used dataset. However, the relatively lower recall reveals the presence of false negatives, i.e., some illicit transactions that were not detected. This limitation may stem from the heterogeneous and intentionally obfuscated behavior of the Lazarus Group, whose operational strategies are designed to conceal transaction patterns and hinder traceability. Consequently, certain transactions may deviate from the dominant behavioral profile learned by the model, leading to occasional misclassifications.
Nevertheless, it is important to remark that, to the best of our knowledge, this is the first evidence that behavioral-based detection in Monero is feasible, highlighting both the promise of the approach and the ample room for further optimization.

\subsection{Results analysis}
While the findings are promising, we recognize that this analysis represents a small-scale case study with inherent limitations. The most significant constraint lies in the very limited support of both the positive and negative classes, reflecting the scarcity of publicly available labeled Monero data. The reduced number of positive samples restricts the model’s ability to learn the full diversity of illicit behaviors and increases the risk of overfitting to the specific operational pattern of the Lazarus Group during the analyzed period. Likewise, a small negative class may not encompass the full range of legitimate user activity, potentially leading to biased decision boundaries. Moreover, the random sampling strategy adopted for selecting licit transactions might overlook other modi operandi—including those that could exhibit similarities to the positive cases—thereby limiting the representativeness of the negative class. In addition, while SMOTE effectively balances the dataset by generating synthetic examples, it cannot fully replace the natural diversity of real-world data. The generated samples are interpolations of existing ones, and therefore may not capture the full complexity of genuine Monero transaction dynamics. Finally, the temporal and contextual scope of the dataset—limited to a single campaign and a narrow time window—further constrains the generalizability of the trained model to other illicit behaviors or future network activity.

\section{Conclusion}\label{sec:conclusions}
This work presents a graph-based approach for analyzing the behavior and modus operandi of criminal users in the Monero blockchain. While the methodology shows promise in uncovering topological patterns linked to illicit activity, this is still a work in progress, and further analysis is required to validate and refine the approach. A key limitation is the dependence on labeled data, which is essential for building and testing the proposed methodology but remains extremely scarce and difficult to obtain. Future work will focus on expanding the dataset to include new behaviors and a more precise form of selecting negative licit transactions.

\section*{Acknowledgment}
This work was partially supported by the European Commission under the Horizon Europe Programme, as part of the projects SAFEHORIZON (Grant Agreement No. 101168562) and FALCON (Grant Agreement No. 101121281).

\bibliographystyle{splncs04}
\bibliography{main}

\end{document}